\documentclass[journal]{IEEEtran}
\ifCLASSINFOpdf
\else
\fi
\usepackage{xspace}
\newcommand{\eg}{\textit{e.g.},\xspace}
\newcommand{\ie}{\textit{i.e.},\xspace}

\usepackage[cmex10]{amsmath}
\usepackage{booktabs}     

\renewenvironment{IEEEbiography}[1]
  {\IEEEbiographynophoto{#1}}
  {\endIEEEbiographynophoto}

\usepackage{comment}
\usepackage{stfloats}
\usepackage{subfigure}
\usepackage{graphicx}
\usepackage{color}
\usepackage[dvipsnames]{xcolor}
\usepackage{placeins}
\usepackage{float}
\usepackage{tabularx,colortbl,multirow}

\usepackage{graphicx}
\graphicspath{{images/}}
\usepackage{enumerate}
\usepackage{enumitem}
\usepackage{amsmath,amssymb,amsfonts,amsthm,tikz,caption}
\usepackage[dvipsnames]{xcolor}
\usepackage{mathtools}
\usepackage{comment}

\usepackage{bm}

\renewcommand\eqref[1]{(\ref{#1})}
\graphicspath{{images/}}
\usepackage{tikz}
	\usetikzlibrary{shapes,decorations,decorations.pathreplacing}
	\usetikzlibrary{arrows,shapes,positioning,calc,fadings,patterns}
	\usetikzlibrary{decorations,decorations.pathreplacing,decorations.pathmorphing}
\usepackage{tikz-3dplot}

\usepackage{booktabs}     
\usepackage{threeparttable}
\usepackage{amssymb}
\usepackage[style=ieee, maxnames=1, minnames=1, backend=biber] {biblatex}
\bibliography{Biblio} 

\makeatletter
\newcommand{\vast}{\bBigg@{4}}
\newcommand{\Vast}{\bBigg@{5}}
\makeatother

\usepackage{siunitx}

\begin{document}
%

\title{Empowering Wireless Network Applications with Deep Learning-based Radio Propagation Models}


%


\author{Stefanos~Bakirtzis,~\IEEEmembership{Member,~IEEE},~Çağkan Yapar,~\IEEEmembership{Member,~IEEE},  Marco~Fiore,~\IEEEmembership{Senior Member,~IEEE},\\ ~Jie Zhang,~\IEEEmembership{Senior Member,~IEEE},~and~Ian~Wassell,~\IEEEmembership{Member,~IEEE} 
\vspace*{-8pt}
}

\markboth{ 
}{Bakirtzis \MakeLowercase{\textit{et al.}}}
\maketitle

\begin{abstract}

The efficient deployment and operation of any wireless communication ecosystem rely on knowledge of the received signal quality over the target coverage area. This knowledge is typically acquired through radio propagation solvers, which however suffer from intrinsic and well-known performance limitations. This article provides a primer on how integrating deep learning and conventional propagation modeling techniques can enhance multiple vital facets of wireless network operation, and yield benefits in terms of efficiency and reliability. By highlighting the pivotal role that the deep learning-based radio propagation models will assume in next-generation wireless networks, we aspire to propel further research in this direction and foster their adoption in additional applications.

\end{abstract}
 
 
\section{Introduction}

\IEEEPARstart{W}{ith} the adoption of new communication technologies, a plethora of challenges emerges in radio propagation modeling due to substantially heterogeneous channel conditions \cite{akyildiz20206g}. Indicatively, and as shown in Fig.~\ref{fig:Wireless_Ecosystem}, next-generation networks are anticipated to support millimeter wave and terahertz communications; these frequency bands are substantially affected by diffuse scattering, whilst the far-field assumption is no longer valid and near-field propagation mechanisms must be considered. Furthermore, installing reconfigurable intelligent surfaces (RISs) will transform the radio propagation environment from a passive medium to an active entity, modulated to enhance communications. In addition, vehicular and high-speed rail communications are dominated by rapidly varying channel conditions, while satellite or unmanned aerial vehicle communications are affected by a set of environmental factors that do not influence legacy cellular networks. 

Consequently, it is becoming increasingly evident that conventional propagation models can no longer meet the pluralistic requirements of next-generation wireless networks. On the one hand,  the traditional approach of trying to accrue and generalize knowledge by fitting mathematical formulas to measured data cannot scale and is unlikely to yield accurate results, due to the complexity of the propagation scenarios.  On the other hand, the full-scale adoption of deterministic site-specific approaches, \eg ray-tracing or finite-difference time-domain (FDTD) simulations, is hindered by the computational resources required to model the propagation environment and emulate the electromagnetic field evolution numerically. In fact, this constitutes a well-known accuracy-efficiency trade-off affecting conventional propagation solvers \cite{Prop_Models_Survey}. As in many other fields, artificial intelligence (AI) emerges as a promising paradigm to cut the Gordian knot and supply radio engineers and mobile network operators with data-driven propagation models that interlace accuracy and computational efficiency. 

That has catapulted substantial research efforts towards developing full-fledged data-driven radio propagation models that will supplant legacy solvers  \cite{AI_Radio_Prop_Comm, overviewRMChallenge}.  Despite having no insight into the mechanisms that affect radio wave propagation, these models harness the profound potential of deep learning  (DL) algorithms to harvest knowledge from data. Their two salient advantages are enhanced accuracy and computational efficiency. Indeed, unlike simplistic empirical and stochastic models, DL-based models employ complex non-linear learnable neural operators that depict the intricate wireless channel behavior. Therefore, they can effectively capture complex dependencies and map properly selected input features to some target quantities of interest (QoIs).  {At the same time, favored by their native graphics processing unit (GPU) implementation, they can emulate wireless channel conditions rapidly via consecutive parallelizable matrix multiplications. 

 
Although there have been detailed reviews on the recent trends in AI-based radio propagation modeling and radio map generation \cite{overviewRMChallenge}, there are no review reports on the benefits stemming from their use in wireless network-related applications. This article fills this gap, underscoring the pivotal role that these models will assume in next-generation wireless networks, fostering their adoption and propelling further research in this direction. To this end, we highlight the potential of AI-based solutions to replace legacy propagation solvers for a set of applications inextricably linked to the successful expansion and operation of the wireless ecosystem. Specifically, we summarize and highlight the distinct advantages arising from their integration into network planning frameworks~\cite{seretis2024fast, bakirtzis2024ai, liu2020cgan}, reconfigurable intelligence surface (RIS)-enabled simulations~\cite{liu2024convolutional, li2022intelligent},  user equipment (UE) localization~\cite{wu2023received, yapar2023real, njima2021indoor, zhao2023nerf2,}, and uncertainty quantification (UQ)~\cite{bakirtzis2022stochastic, seretis2019uncertainty}. These use cases effectively exemplify the capacity of DL-based models to streamline various essential network operations, inaugurating a new era of automation in the wireless ecosystem, and paving the way for the adoption of these models in new applications.

 \begin{figure*}
\centering
    {\includegraphics[width = 1\textwidth]{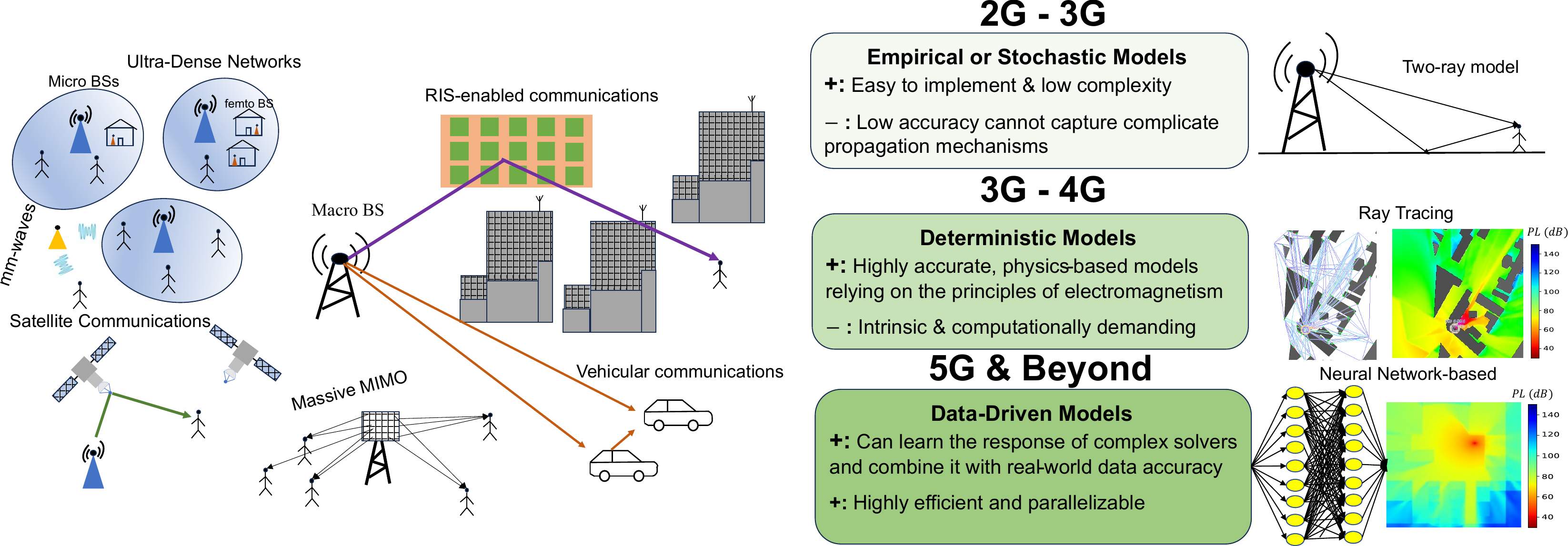}}
\caption{{Next-generation wireless ecosystem comprising a plethora of diverse wireless channels, each posing unique challenges and affected by different propagation mechanisms; radio propagation models have been evolving alongside to cope with these  challenges }}


\label{fig:Wireless_Ecosystem}
\vspace*{-12pt}
\end{figure*}

 \section{Radio Propagation Modeling Under the Prism of Deep Learning}

Radio propagation models are interwoven with the efficient and successful design of any wireless communication system. Having reliable estimates of the spatiotemporal wireless signal distribution allows the proactive network design so as to attain the highest signal-to-interference-plus-noise ratios. In turn, this enables the use of appropriate modulation and coding schemes that yield higher data rates. The following two subsections introduce conventional radio propagation modeling and then focus on DL-based techniques, respectively.

\subsection{Conventional Propagation Models}

To model radio wave propagation, one should either observe the signal quality of existing wireless communication networks or have laws that govern the propagation of electromagnetic waves. Empirical or statistical models follow the first approach, furnishing unsophisticated mathematical models whose parameters are fine-tuned through field data. Site-specific deterministic models hinge on Maxwell's equations that constitute the foundations of electromagnetic theory. However, even in the simplest propagation environments, Maxwell's equations are analytically intractable, though numerical solutions are feasible when appropriate approximations are made. For instance,  ray-tracing approximates the waves as rays and computes their phase and amplitude through the eikonal and transport equation. Another example is the FDTD technique, which employs a centered finite-difference to approximate the partial spatial and temporal equations in Maxwell's equations and attain an iterative transient solution. Such approaches deliver much more reliable results, but at the same time are computationally demanding, thus provoking the accuracy-efficiency dilemma mentioned previously.

\subsection{Deep Learning Data-driven Propagation Models}

The advent of AI and DL   propelled a paradigm shift in radio propagation modeling, aiming at overcoming the inherent accuracy-efficiency trade-off of conventional propagation modeling techniques, via learning-based neural approaches.  Figure~\ref{fig:Wireless_Ecosystem} exemplifies the utilization of simplistic empirical or semi-empirical models in first-generation wireless networks, such as the two-ray model, evolving into more sophisticated propagation models in subsequent generations, tracing thousands of rays to compute the received signal. We envision that in fifth-generation and beyond (5B/B5G) the rays will be gradually replaced by learnable links between neurons that assimilate radio propagation characteristics. 

\begin{figure*}
\centering
    {\includegraphics[width = 0.95\textwidth]{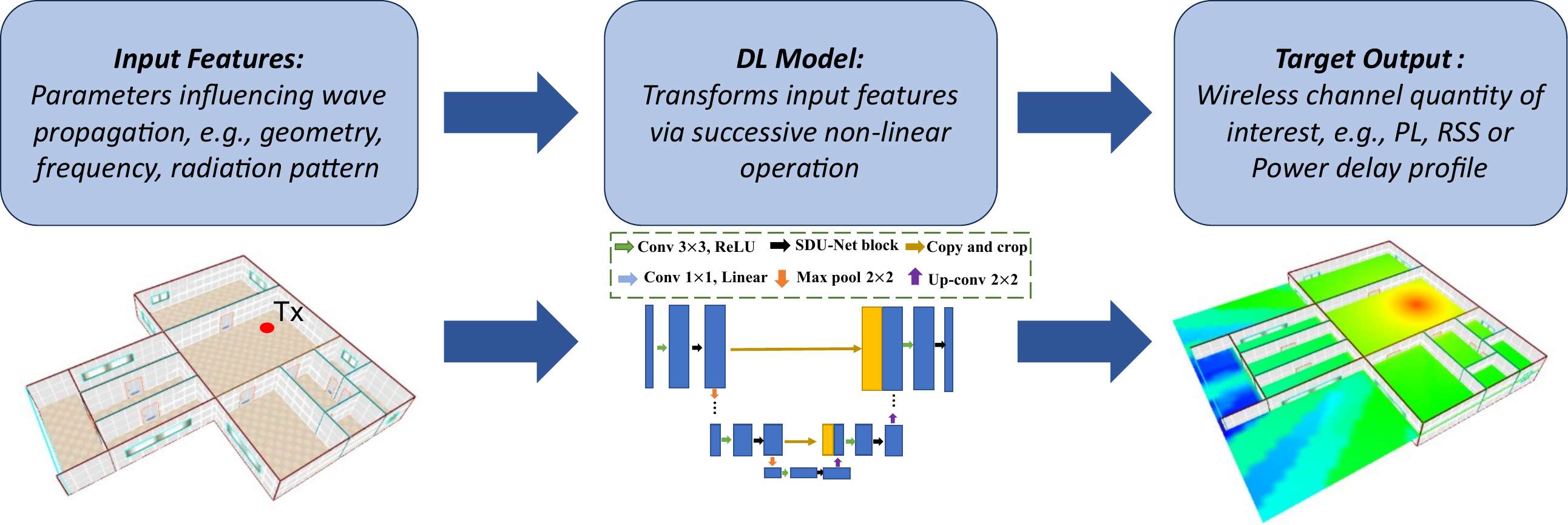}}
\caption{{Characteristic pipeline of a data-driven propagation model: a set of input features are processed by a DL module which transforms it to a wireless channel QoI.}}

\label{fig:Data_Driven_Pipeline}
\vspace*{-12pt}
\end{figure*}

 The general pipeline of a data-driven propagation model is depicted in Fig.~\ref{fig:Data_Driven_Pipeline}. As can be seen, it comprises three constituent components: (i) a set of input features, (ii) a DL model, (iii) and some target outputs.  The framework's goal is to tune the learnable parameters of the DL model to effectively transform the input data into a QoI related to the wireless channel. The selected input features assume a decisive role in the success of the learning process and the generalizability of the model, hence they must convey information that is intrinsically linked to the parameters that affect the target QoI. These parameters can be environmental or related to the wireless network configurations. Examples of environmental parameters that affect radio propagation are the city layout, the building height, and the terrain type for outdoor environments or the building layout, and the construction material type for indoor environments. The wireless system parameters can be related to the position of the transmitter (Tx), the antenna radiation pattern, and the transmitting frequency.

These pieces of information will be then processed by a DL model. Radio propagation modeling is typically treated as a supervised learning problem, and the DL model is used to identify a causal connection and mapping between the input and output data. Commonly used DL architectures used for this purpose are multilayer perceptions (MLPs), convolutional neural networks (CNNs), and more recently, transformers. 
In MPLs, the learnable components (neurons) are arranged into interconnected layers, and, starting from the input, they apply non-linear transformations to the weighted sum of the previous layer information. In CNNs, learnable filters/kernels are used to process the input data, capture local patterns, and extract new high and low-level features that can be subsequently translated to the target QoI. Transformers exploits the concept of multi-head self-attention, indicating how much focus the model should attribute to each input element,  implemented through some learnable query, key, and value vectors.

The output QoIs that the DL model aims at inferring can be related to large or small-scale fading attributes, \eg path loss (PL), received signal strength (RSS), shadowing, angle of arrival (AoA), or the power-delay profile. Since most of these physical quantities are non-quantized, radio propagation modeling is usually posed as a regression problem.  The   QoI values can be collected through extensive channel measurement campaigns or they can be generated in simulation environments using high-performance radio propagation models. The use of high-fidelity simulation data or real-world data ensures the accuracy of the data-driven model, while their architectural simplicity and straightforward parallelization guarantee their computational efficiency.  

The data used to implement an AI-driven propagation model are separated into three different datasets: a training, a validation, and a test dataset. The first acts as a source of knowledge for the DL model utilized to estimate its learnable weights, such that the model's predictions mimic the target ground truth.  Commonly used error metrics used to assess the prediction quality are the root mean square error (RMSE), the mean absolute error (MAE),  or the mean absolute percentage error (MAPE).  The validation data are used to adjust the DL model hyperparameters and select the model weights instance that attains the best performance, \ie minimizes the error metric. Finally, the test data are used to evaluate the performance of the data-driven framework to unknown  (not seen during training) propagation scenarios, \eg new urban environments or different frequency bands, and assess its generalizability.

 \vspace{-2mm}

\section{Applications}

In the following sections, we discuss how data-driven propagation models can streamline important wireless network applications, such as those shown in Fig.~\ref{fig:Applications}, including network planning, RIS-enabled system simulation, localization, and UQ. For compactness, details of this discussion are also summarized in Table~\ref{table:Overview}.

\subsection{Network Planning}

Currently, the de facto approach in network planning relies on coupling AI optimizers with radio propagation solvers to smartly and expediently scan the optimization space of possible network deployments. Specifically, the optimization algorithms guide the Tx site selection, while improving some key performance indicators (KPIs) evaluated based on wireless channel estimates computed via radio propagation models, \eg a ray-tracer. As the optimization problem is highly dependent on the target deployment environment, typically these problems do not have a closed-form solution, therefore, planning frameworks commonly employ heuristic optimization algorithms. 

The performance of such network planning frameworks is highly dependent  on the radio propagation solvers used since ($i$)~the KPI fidelity relies on the accuracy of the solver, and ($ii$)~the optimization time is dominated by the time required for the solver to emulate channel conditions. In turn, this relates to the accuracy-efficiency trade-off of conventional propagation models, mentioned in the introduction,
and motivated the works in~\cite{seretis2024fast, bakirtzis2024ai} to explore how DL-based propagation models can expedite the planning process without compromising its effectiveness. An initial implementation in this direction is to utilize the radio maps from some candidate Tx locations in a propagation environment to train a data-driven model. Once trained using only a limited number of samples, the DL-based model can be employed during the optimization process to generate radio maps for any Tx position in the geometry,  substituting the ray-tracer.  This method was exemplified in \cite{seretis2024fast}
where a U-Net-like convolutional encoder-decoder (CED)  received as input some physics-based information and predicted the RSS for each Tx.  For $89 \%$ of the optimization runs the optimization framework identified a deployment that met the coverage and capacity requirements while yielding a $10 \times$ reduction in the computational runtime.

\begin{table*}[t!]
  \begin{center}
    \caption{\sc{summary of works integrating  DL-based propagation models into wireless network applications.}}
    \label{table:Overview}
 
   \resizebox{2\columnwidth}{!} {\begin{tabular}{|c    ||  c || c || c  ||  c   |  }
\hline

 { \cellcolor{blue!15}  \textbf{Application}  }   &     { \cellcolor{purple!15}  \textbf{Work}  } &  \cellcolor{brown!15}  DL Model &  {  \cellcolor{yellow!15} \textbf{Summary}  } & %
     { \cellcolor{green!15} \textbf{Benefits}  }      \\

\cellcolor{blue!15}  &    \cellcolor{purple!15}   {  }  & \cellcolor{brown!15}  & \cellcolor{yellow!15}  &  \cellcolor{green!15}      \\

\hline
  &     \cite{seretis2024fast}     &  CED & \begin{tabular}{@{}c@{}} Conducted few ray-tracing simulation \\  to train CED and use it in optimization\end{tabular}    &  \begin{tabular}{@{}c@{}} 10$\times$ reduction \\ in planning time \end{tabular}        \\
    
    \cline{2-5}                 
    \begin{tabular}{@{}c@{}} Network  \\  Planning \end{tabular}   &    \cite{bakirtzis2024ai}  &  CED   &                     \begin{tabular}{@{}c@{}} Replaced ray-tracer with fully- \\ fledged \textit{pre-trained} DL-based model \end{tabular}     &   \begin{tabular}{@{}c@{}} 100$\times$ reduction \\ in planning time \end{tabular}  \\

    \cline{2-5}
    &   \cite{liu2020cgan}   & CED &  \begin{tabular}{@{}c@{}} Trained a CED to directly infer \\  the heatmap for optimal topology \end{tabular}      & \begin{tabular}{@{}c@{}} Single Inferrence $ \leq  $ 1 s\\  planning approach \end{tabular} \\
                     
    \hline
    \hline
    
  \begin{tabular}{@{}c@{}}    \\  RIS \end{tabular}   &     \cite{liu2024convolutional}  & CNN-MLP & \begin{tabular}{@{}c@{}} Inferred the \textit{simulated} near $\&$ far field  scattering\\   patterns based on RIS binary configuration \end{tabular}     &   \begin{tabular}{@{}c@{}}   Infers RIS pattern using \\  $ < 1\%$ of RIS configurations \end{tabular}     \\

       \cline{2-5}
         Modeling        &   \cite{li2022intelligent}  & \begin{tabular}{@{}c@{}}   PINN \\ with MLPs \end{tabular}   &   \begin{tabular}{@{}c@{}} Forward and inverse configuration  to   \\ \textit{measured}  RIS scattering pattern mapping   \end{tabular}  &  \begin{tabular}{@{}c@{}}   Real-time realistic RIS pattern  \\   or configuration inferrence \end{tabular}       \\

    \hline  
    \hline
    
\multirow{4}{*}{Localization}  &  \cite{wu2023received}   &    GAN &  \begin{tabular}{@{}c@{}}  On-the-fly fingerprint radio map generation \\ by probing \textit{simulated} RSS distribution \end{tabular}  &    \begin{tabular}{@{}c@{}}   $40 \times$  time reduction in \\    fingerprinting map generation  \end{tabular}   \\
   
    \cline{2-5}
    &    \cite{yapar2023real}  & CED &  \begin{tabular}{@{}c@{}}  Inferring probability a UE being in grid location \\  via DL-based generated radio maps \end{tabular}    &  \begin{tabular}{@{}c@{}} Fast positioning resilient \\  to radio map inaccuracies  \end{tabular}    \\
   \cline{2-5}
    &   \cite{njima2021indoor}  & GAN &     \begin{tabular}{@{}c@{}} Probed limited volume of RSS \textit{measurements}   \\   and generated additional high-fidelity data \end{tabular}  &    \begin{tabular}{@{}c@{}}   $15 \% \uparrow $   localization accuracy \\   via realistic data augmentation  \end{tabular}    \\
    \cline{2-5}
    &    \cite{zhao2023nerf2}  & MLPs &  \begin{tabular}{@{}c@{}}   Neural ray-tracer inferring spatial   \\  spectrum based on \textit{measured data}\end{tabular}     &  \begin{tabular}{@{}c@{}}   Median error $ \downarrow 50\%$   \\ and  error variance
  $ \downarrow  40\%$ \end{tabular}        \\

    \hline
    \hline
    
 \begin{tabular}{@{}c@{}}    \\  Uncertainty \end{tabular}    &   \cite{bakirtzis2022stochastic} &  CED &   \begin{tabular}{@{}c@{}} Replaced ray-tracer with fully- \\ fledged \textit{pre-trained} DL-based model \end{tabular}     &   \begin{tabular}{@{}c@{}}   $\sim 100\times$ reduction \\ in UQ time \end{tabular}     \\
    \cline{2-5}
      Quantification   &    \cite{seretis2019uncertainty}  & MLP &  \begin{tabular}{@{}c@{}} Inferred signal attenuation over distance   \\  according to underground tunnel characteristics \end{tabular}   &    \begin{tabular}{@{}c@{}} 20$\times$ acceleration \\ in UQ convergence \end{tabular}    \\

    \hline

      \multicolumn{5}{|c|} { \begin{tabular}{@{}c@{}}    CED: Convolutional Encoder-Decoder, CNN-MLP: Convolutional Neural Network terminated with Multilayer Perceptron \\  PINN: Physics Informed Neural Network, GAN: Generative Adversarial Network \end{tabular}  }  \\ 

     \hline

\end{tabular} }
\end{center}

\end{table*}

 Instead of creating new samples for each optimization geometry under consideration, a more general strategy is to develop a fully-fledged data-driven propagation model, by training a  DL architecture in advance with a large volume of data from diverse complexity propagation environments and network configurations. This approach was presented in \cite{bakirtzis2024ai}, where a CED-based model, called EM DeepRay, was pre-trained to transform physics-based information (electromagnetic material properties, physical distance, free space PL) into the PL distribution for numerous indoor environments and frequency bands. Once trained, the model could reliably infer the PL distribution for new environments, not seen during training,  and consequently evaluate the attained throughput throughout the building. After exploring different optimization metaheuristic algorithms, \eg genetic algorithms and particle swarm optimization,  it was shown that if one conducts the optimization using a ray-tracer or their standalone data-driven propagation model they should expect to find a topology that attains approximately the same performance and is equally good. At the same time,  fully eliminating the use of the ray-tracer, decreased the optimization runtime by two to three orders of magnitude. 

A different and more rigorous approach is to translate network design into an end-to-end single-step inference process. In this case, a DL architecture learns to observe a propagation environment and directly infers the RSS distribution for the best network topology. To achieve this,  a dimension-aware conditional generative adversarial network (DA-cGAN) was employed in \cite{liu2020cgan}, receiving as input a colored representation of the indoor floor plan,  features related to the building dimensions, and a heatmap of the interference introduced by the macro BS. At its output, DA-cGAN predicted directly the optimal position of the Txs throughout the indoor geometry along with the equivalent RSS distribution. To train the proposed framework, various actual floor plans from internal Ericsson customers were used, and then highly experienced radio engineers determined the positions where the Txs were to be deployed and generated the radio maps with an in-house radio tool. Ultimately, the planning process amounted to a negligible time of less than 1 second, and DA-cGAN pronounced on average the same number of Txs as the ground truth deployment, with a median displacement of 3 m.

\subsection{Reconfigurable Intelligence Surface Modeling}

The integration of RISs into the wireless ecosystem aspires to revolutionize the operation of legacy communication systems and establish a next-generation of highly intelligent, automated, and energy-friendly wireless networks. Indeed, instead of deploying new Txs and emitting new signals, RISs can redirect electromagnetic waves and steer them in specific directions, thus augmenting network sustainability, enhancing the received signal, and improving the quality of service. To support realistic RIS-enabled wireless network simulations,  accurate and fast models that can map the RIS configuration to its radiation pattern, or vice-versa, are required to precisely emulate channel conditions. That is typically accomplished by either conducting measurements \cite{li2022intelligent} or through full-wave simulations with dedicated tools, such as the HFSS finite-element software \cite{liu2024convolutional}. 
However, the possible RIS configurations increase exponentially with the RIS size; so, for a RIS with  $M \times N$ cells and  $c$  states per cell, the total number of configurations is equal to $c^{M \cdot N}$. For instance, a $4 \times 4$ cell RIS with binary cell states,  gives a total of  $2^{16}$ configurations. Evidently,  it is not feasible to conduct measurements for all the possible configurations and full-wave simulations are extremely time-consuming requiring more than 1 hour per RIS configuration \cite{li2022intelligent}.

 \begin{figure*}
\centering
    {\includegraphics[width = 1\textwidth]{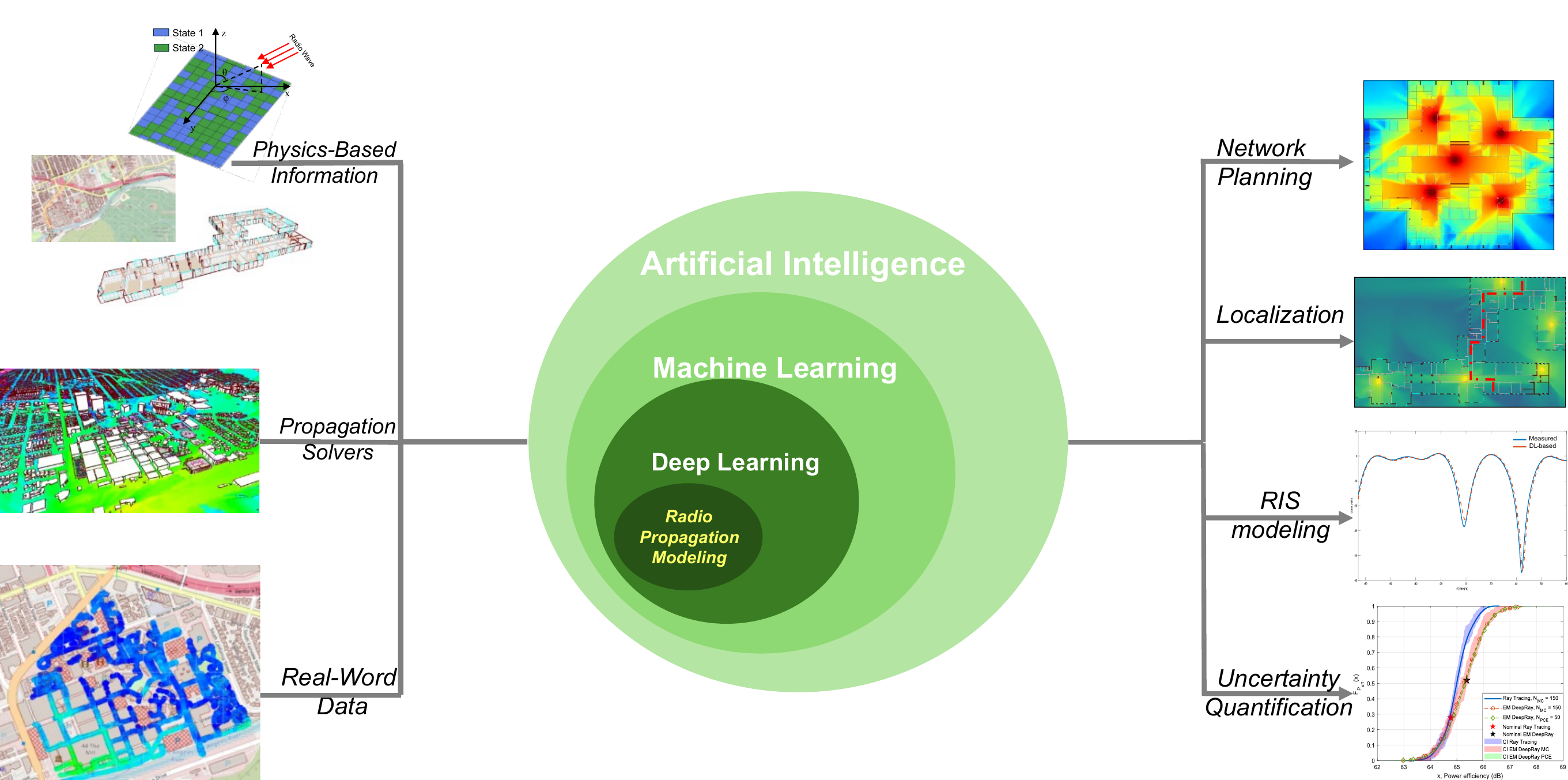}}
\caption{{Visualization of the exploitation of DL-based model for empowering wireless network applications.   }}

\label{fig:Applications}
\vspace*{-12pt}
\end{figure*}

 Recent works have shown that data-driven approaches can overcome these limitations \cite{liu2024convolutional, li2022intelligent}, effectively replicating full-wave analysis or measurement data.  The first was presented in \cite{liu2024convolutional}, where a RIS with binary cell configurations was considered, and four steps were followed to generate data. Initially,  two full-wave simulations were conducted, with all cells configured at state 0 or  1 to compute the tangential electric fields over each cell at the RIS surface. Subsequently,  some cell states were randomly changed, and it was assumed that the fields for these cells would correspond to those computed in the first step. For instance, if the state of the (i,j) cell was 0 they assigned to its tangential electric fields the values that the cell fields had in the first step when its state was 0 (note that now not all cells had the same state). In the third step,  full-wave simulations with HFSS were run to calculate the actual field response, and finally,  the difference between the simulated values and the values assumed in the second step was computed. Then, a CNN  received as input an $N \times M$ tensor depicting the cell configuration and was trained to infer the difference in the real and imaginary part of the tangential electric field computed in the third step. Ultimately, the final tangential fields were found by adding the inferred difference to the values assumed in the second step, and it was shown that the proposed framework could accurately infer both the near and the far-field scattering RIS patterns for unknown RIS configurations not seen during the training phase. In addition, when plugging the inferred RIS pattern into a ray-tracer to conduct link budget analysis, the authors observed results similar to those derived using the full-wave solver. Overall, the proposed approach required simulating less than $1\%$ of the possible RIS configurations,   rendering it feasible to emulate the RIS near and far-field pattern for any cell configuration.

Instead of using simulated data, measurements can be harnessed to enhance the fidelity of the model, while it is also possible to map patterns to RIS configurations to underpin beamforming techniques. To achieve that,  in \cite{li2022intelligent} a smartly designed automeasuring system was used,  recording the radiation pattern of the RIS for $25\%$ of the possible RIS configurations.  Then,  a physics-informed neural network (PINN) inspired by the discrete dipole approximation (DDA) algorithm was introduced. Specifically, the weights of the PINN corresponded to the matrix used in DDA to calculate the total magnetic field. The input of the PINN was the RIS configuration multiplied by the real and imaginary incident magnetic field, and the weights were updated iteratively via backpropagation until the output mimicked the desired radiation pattern. To assess the few-shot learning potential of the proposed framework, the PINN was trained with $20 - 80 \%$ of the sampled data (corresponding to $5-20\%$ of the possible configurations) and the corresponding MSE was between 2.2 to 2.5 dB. This result corroborated that the PINN can learn with a limited volume of data and can accurately and efficiently infer the scattering pattern. Consequently, the PINN was used to generate a larger volume of data used to train a 7-layer MLP to solve the inverse problem with a $99 \%$ accuracy, \ie from the radiation pattern to find the corresponding configuration. The latter was exploited to support an on-the-fly intelligent beamforming system that allowed the RIS configuration to be rapidly inferred based on the desired radiation pattern.

 
\subsection{Localization}



 
 Cellular and WiFi networks have been considered as alternative localization technologies in scenarios where satellite-based location tracking accuracy degrades, \eg in dense urban or indoor environments dominated by non-line-of-sight conditions. In such cases, the UE  location is identified via the characteristics of the wireless signals, \eg RSS, or AoA. A common approach is fingerprinting, which, as the name implies,  relies on matching a measured feature from multiple Txs at the UE side (commonly the RSS)  to a pre-recorded measurement database of the same feature within a certain area. Fingerprinting comprises two phases: (i) the offline phase, during which the database is built by collecting measurements, and (ii) the online phase at which a new ``fingerprint" is sampled and compared to the ``fingerprint database".

The fingerprint databases can be built either through measurement campaigns or propagation models.  Constructing databases through measurement campaigns is a laborious and time-consuming process, while the databases can also become outdated as the propagation environment or the wireless system configuration changes, \eg addition of new objects in the propagation environment or changes in the antenna beamforming radiation pattern. Therefore, generating simulation-based databases with reliable accurate propagation models is a much more viable and commonly selected approach. 


Data-driven propagation models can seamlessly support the generation of fingerprinting databases and relevant meta-features. For instance, RSS fingerprinting radio maps were generated in \cite{wu2023received} via a generative adversarial network (GAN). Specifically, the generator received as input a map of the geometry and the Tx position and was tasked to generate the respective radio map.  At the same time, a discriminator was responsible for discerning whether the synthetically generated radio maps were real or fake, \ie whether they were similar to the ray-tracing ground truth fingerprint map. Ultimately, the generator was trained to produce such realistic radio maps that it was possible to deceive the discriminator which could no longer distinguish between the generated and ground truth fingerprint maps. Using these maps for localization the framework attained a $92.15\%$ pixel-level accuracy while reducing the average runtime by approximately 70 times.   Meta-features databases for localization in outdoor environments were created in \cite{yapar2023real}, where the authors introduced LocUNet. Assuming a set of $N_{BSs}$ BSs, LocUNet received as inputs (i) the RSS values of each BS reported by the UE,  (ii)  radio maps for each BS, generated by another U-Net and depicting the RSS distribution throughout the city, (iii) a map of the urban environments, (iv) the position of all the BSs. Then, LocUNet outputted a map portraying the likelihood that the UE is located at each map point. Through a set of exhaustive comparisons, it was shown that the proposed framework outperformed other well-established localization algorithms in realistic urban scenarios, and it yielded high robustness to inaccuracies in the radio map estimates.

In addition, DL can be leveraged to synthetically augment the volume of measured data. This paradigm was presented in \cite{njima2021indoor} using a GAN. Specifically, the generator was exposed to some real-world RSS indicator (RSSI) fingerprinting measured data and managed to comprehend the underlying RSSI distribution. Consequently, the generator was employed to produce high-fidelity data according to this distribution, enabling an artificial increase in the dataset size and diversity, as well as,   diminishing the reliance on massive time-consuming measurements. Then, an MLP was trained to infer the UE location based on the original and the synthetically generated RSSI.  The proposed framework, exploiting the artificially generated data,  achieved up to a $15\%$ improvement in localization accuracy.
 
 Instead of simply generating RSS heatmaps, a more holistic approach is to design DL-based models that can replicate the entire spatial spectrum at the UE, \ie emulate the strength of the multipath RF signals arriving from various azimuth and elevation angles. A characteristic example of this category is the neural ray-tracing module presented in \cite{zhao2023nerf2} referred to as $NeRF^2$.  A huge dataset of measured data was collected at some 500,000 positions from 14 indoor scenes. Using only $10 \%$  of the data, through a turbo learning approach, $NeRF^2$ could infer the spatial spectrum with a structural similarity index equal to 0.82, almost 2.5 times better than ray-tracing, and the AoA median error was only $1.96^{o}$.   Then, employing $NeRF^2$ for localization in a real setting of an elderly nursing home with 50 Bluetooth low energy  gateways,  a precision of 1.41 m in the UE positioning was attained, yielding an almost $50 \%$ improvement compared to that with ray-tracing.

\vspace{2mm}
 
\subsection{Uncertainty Quantification}


 Wireless channel characteristics are intrinsically affected by multiple non-deterministic parameters.  Hence, to improve network reliability and reduce relevant risks, the impact of the environmental input uncertainty introduced into the wireless channel characteristics ought to be quantified. To achieve this, one can treat the propagation parameters as random variables following a certain probability density function.  These random parameters can refer to either the physical properties of the environment or its geometric characteristics. Then,  the stochastic channel characteristics can be evaluated repeatedly through a propagation solver, probing the propagation environment parameters from the corresponding sampling distributions, resulting in a probability distribution of the channel QoI.

  Similar to network planning, UQ can accomplished via a standalone pre-trained data-driven propagation model or by training a DL algorithm for a specific propagation scenario to learn from a limited number of observations the impact of randomness on a QoI.  An example of the first case was shown in  \cite{bakirtzis2022stochastic}, where EM DeepRay was employed to quantify the uncertainty in PL and throughput due to the randomness of electric permittivity and conductivity of the walls.  The PL and throughput probability distributions for  LoS and nLoS points were evaluated through ray-tracing and EM DeepRay, yielding a close resemblance. Yet, the UQ analysis with the data-driven propagation model required less than a minute, while via ray-tracing took a couple of hours. Interestingly, both models indicated the impact of uncertainty was more pronounced for nLoS points. A case falling in the second category was studied in \cite{seretis2019uncertainty} for a railway communication system assuming uncertainties in the tunnel geometry.  The propagation of electromagnetic waves through underground arched tunnels was emulated via a vector parabolic equation solver, and an MLP was trained to infer the RSS across the tunnel as a function of the tunnel radius, the receiver lateral position, and height to the center distance. The MLP required approximately 100 simulations to comprehend the impact of the geometry parameters on the RSS output distribution, while the conventional Monte Carlo method required almost 10 times more simulations to converge.




\section{Open Challenges and Future Directions}

\textbf{Open-access reference datasets.} Currently, the preponderance of the existing standalone data-driven propagation models builds on synthetic data extracted from asymptotic propagation solvers, \eg ray-tracers. Yet, these models intrinsically exhibit errors with respect to the real channel behavior due to the finite accuracy of the propagation tool, \eg approximation of electromagnetic wave with rays, the way that the digital environment is simulated (wall thickness, material properties), and details that cannot be captured in the simulation (small objects and human motion). This calls for open-access databases of measured signal data, that could be exploited in conjunction with simulated data, to train DL models that probe the underlying distributions of both data modalities, amending the innate fidelity limitations of a propagation solver. This is an important capability that conventional solvers do not have, \ie measurements can be only used to tune their hyperparameters or adapt their predictions on a per-scenario basis. However, DL-based approaches can assimilate this information to holistically improve their performance by adapting the  signal distribution produced by the model \cite{overviewRMChallenge}.  

\textbf{Learning ray trajectories.} Another interesting direction would be to forge data-driven models that learn tracing rays per se, rather than simply inferring a channel characteristic. Indeed, existing data-driven propagation models focus on inferring directly the results of a legacy solver, \eg PL or RSS, at certain points of interest.  Instead, one could train models that learn to replicate the full solution of a solver, \eg the entire ray trajectory, leading to increased accuracy and allowing various correlated measures, such as multipath delay, and angles of arrival/departure to be inferred simultaneously.   
 
\textbf{LLMs for radio propagation.} Finally, we believe that a noteworthy direction would be to explore the use of large language models (LLMs) in radio propagation modeling. These models have demonstrated a unique potential to process large datasets of images, identify patterns and correlations within them, and consequently create new realistic images with remarkable faithfulness. Hence, one could perceive radio map generation as an image generation problem, and expose an LLM to images depicting radio propagation environments and the respective radio map images. Additional information related to the nature of radio propagation could also be contextualized and provided to the LLM in the form of prompts, \eg "When electromagnetic waves impinge onto walls you should observe a reflected and a transmitted component".

\section{Conclusion}

 This article reviewed and highlighted the benefits stemming from the use of data-driven propagation models in some critical network applications. These include efficient network planning strategies, rigorous RIS response modeling, robust UE localization schemes and reliable UQ. We anticipate that in the future, DL-based propagation models will be adopted in more applications, and will be key enablers for real-time network automation.

\section*{Acknowledgment}

The work of Stefanos Bakirtzis is supported by the Onassis Foundation and
the Foundation for Education and European Culture.

 \printbibliography

\end{document}